\documentclass[secnumarabic,amssymb,nobibnotes,aps,prd]{revtex4}
\usepackage[dvipdfm]{graphicx}
\usepackage{latexsym,mathrsfs}
\usepackage{dcolumn}
\usepackage{bm}

 \newcommand{\iu}{\mbox{i}}
\newcommand{\is}{\mbox{\scriptsize i}} 
\newcommand{\aq}{& = &}

\def \av#1{\langle #1\rangle}

\def \ve#1{\mbox{\boldmath $#1$}}

\def\dddot#1{\mathop{%
	\ooalign{{\hss\hbox to.8ex{$\dot{}\hss\dot{}\hss\dot{}$}\hss}\crcr{$\hss#1\hss$}}}}

\makeatletter
\def\iddots{\mathinner{\mkern1mu\raise\p@
    \hbox{.}\mkern2mu\raise4\p@\hbox{.}\mkern2mu
    \raise7\p@\vbox{\kern7\p@\hbox{.}}\mkern1mu}}

\begin{document}

\title{Asymptotic behavior of the mean square displacement of the Brownian parametric oscillator near the singular point}

\author{Tohru Tashiro}
\affiliation{Department of Physics, Ochanomizu University, 2-1-1 Ohtuka, Bunkyo, Tokyo 112-8610, Japan}

\date{\today}

\begin{abstract}
A parametric oscillator with damping driven by white noise is studied. The mean square displacement (MSD) in the long-time limit is derived analytically for the case that the static force vanishes, which was not treated in the past work \cite{tashiro07}. The formula is asymptotic but is applicable to a general periodic function. On the basis of this formula, some periodic functions reducing MSD remarkably are proposed.
\end{abstract}

\maketitle

\section{Introduction}

We shall deal with the Brownian motion described by the following Langevin equation which is the same one in the previous work \cite{tashiro07}:
\begin{equation}
\ddot{x}(t) + \beta\dot{x}(t) + \left[w+q\phi(t)\right]x(t) = f(t)
\label{eq1}
\end{equation}
where $x(t)$ is a position of the Brownian particle at time $t$, $\beta$ is a damping constant per unit mass and $f(t)$ is a centered Gaussian-white noise with correlation function
\begin{equation}
\av{f(t)f(t')} = 2\epsilon\delta(t-t') \ .
\end{equation}
$\phi(t)$ is a periodic function of $\pi$ whose maximum and minimum values are $\pm 1$ as defined in Ref.\cite{tashiro07}. Even if the period is not $\pi$, we can obtain Eq.(\ref{eq1}) again by introducing scaled parameters. In the motion of a charged particle in a Paul trap with buffer gas near standard temperature and pressure by Arnold et al. \cite{arnold93}, which is accurately modeled by this Brownian motion, $w$ and $q$ in Eq.(\ref{eq1}) correspond to a dc and an ac voltage, respectively.

It is reported in the previous paper \cite{tashiro07} that for a general periodic function $\phi(t)$, three variances in the long-time limit under the condition $\av{x(t)}=\av{v(t)}=0$ \footnote{As one can see, $\av{x}$ and $\av{v}$ are governed by the deterministic equations, $\partial_t{\av{x}}=\av{v}$ and $\partial_t{\av{v}} + \beta\av{v} + \left[w+q\phi(t)\right]\av{x}=0$. Thus, these conditions are equivalent to $x=v=0$ at initial time. Even if the initial conditions are not assumed, both $\av{x}$ and $\av{v}$ go to 0 in the long-time limit since we shall suppose that parameters which make this damped parametric oscillator stable in the long-time limit are selected.}, 
\begin{equation}
\sigma_{x^2}(t) \equiv \av{x^2(t)} \ , \ \ \sigma_{v^2}(t) \equiv \av{v^2(t)} \ \ \mbox{and} \ \ \sigma_{xv}(t) \equiv \av{x(t)v(t)} \ ,
\label{vars}
\end{equation}
can be obtained analytically by using series of $q$ as long as $w>0$ holds. Moreover, it is showed that as $q$ increases the mean square displacement (MSD) can be suppressed less than that at $q=0$, i.e., MSD of a Brownian motion in a harmonic potential. We called this phenomenon a {\em classical fluctuation squeezing}.

In the experiment with the Paul trap, a static force produced by a dc voltage balances with gravity, which means that we should set $w=0$ in Eq.(\ref{eq1}). 
For this case with $w=0$, however, MSD in the long-time limit cannot be represented by a series of $q$ in \cite{tashiro07},
\begin{equation}
\overline{\sigma_{x^2}} = \frac{\epsilon}{\beta w} + \sum_{n=1}^{\infty}a_nq^n \ ,
\end{equation}
since the zeroth-order term diverges, which means that $(w,q)=(0,0)$ is a singular point of MSD in the long-time limit. We can understand this from another point of view. Equation (\ref{eq1}) with $w=q=0$ denotes a free Brownian motion. Therefore, MSD in the long-time limit goes to $\infty$ because of a well-known relation, $\sigma_{x^2}(t)\propto t$. In short, the theory about MSD in the previous paper is right only for the case with $w>0$ and cannot be used at $w=0$. Thus a new method is desired in order to solve the system with $w=0$.

In this paper we will investigate the asymptotic behavior of MSD in the long-time limit near the singular point $w=q=0$ for general periodic functions, which is not taken up in the previous work. Of course, investigations for this case have been done for the Mathieu's case of $\phi(t) = 2\sin2t$ \cite{arnold93,hanggi},  but there is no study dealing with the system with general periodic functions as far as we know. In fact, we will show that MSD for other periodic functions can be smaller than that for the Mathieu's case. This result will reveal a significance of employing other periodic functions in experiments using the Paul trap with a buffer gas.

\section{Periodicity of variances in the long-time limit}
\label{period}

Here, we shall examine the periodicity of variances in the long-time limit which is a very important property when investigating the asymptotic behavior of MSD. Of course, the periodicity is clear in the previous paper \cite{tashiro07}, but the proof is valid as long as $w > 0$ holds. The proof on this section remains true even if $w\le0$.

Three variances fulfill the following differential equation 
\begin{equation}
\frac{d}{d t}\ve{X}(t) = -\left[\ve{W}+q\ve{\Phi}(t)\right]\ve{X}(t) + \ve{K} \ ,
\label{eq2}
\end{equation}
with
\begin{equation}
\ve{X}(t) \equiv \left(
\begin{array}{c}
\sigma_{x^2}(t) \\
\sigma_{v^2}(t) \\
\sigma_{xv}(t)
\end{array}
\right) \ , \ \ \ve{W} \equiv \left(
\begin{array}{ccc}
0 & 0      & -2 \\
0 & 2\beta & 2w \\
w & -1     & \beta
\end{array}
\right) \ , \ \ \ve{\Phi}(t) \equiv \phi(t)\ve{P} \equiv \phi(t)\left(
\begin{array}{ccc}
0 & 0 & 0 \\
0 & 0 & 2 \\
1 & 0 & 0
\end{array}
\right) \ {\rm and} \ \ve{K} \equiv \left(
\begin{array}{c}
0 \\
2\epsilon \\
0
\end{array}
\right) \ .
\end{equation}
(See appendix \ref{app3}.)
For simplicity, we represent the coefficient matrix by $\ve{A}(t)$:
\begin{equation}
\ve{A}(t) \equiv -\left[\ve{W}+q\ve{\Phi}(t)\right] \ .
\end{equation}

It is well known that a solution of Eq.(\ref{eq2}) can be obtained by using 
the Green's function of the homogeneous equation like this:
\begin{equation}
\ve{X}(t) = \ve{G}(t,t_0)\ve{X}(t_0) + \int_{t_0}^{t}d t_1\ve{G}(t,t_1)\ve{K}
\label{eq3}
\end{equation}
where
\begin{equation}
\ve{G}(t,t_0) \equiv \ve{1} + \int_{t_0}^{t}d t_1\ve{A}(t_1) + \int_{t_0}^{t}d t_1\int_{t_0}^{t_1}d t_2\ve{A}(t_1)\ve{A}(t_2) + \cdots \ (t \ge t_0)
\end{equation}
in which $\ve{1}$ is the 3 $\times$ 3 unit matrix.

One can easily confirm that $\ve{G}$ has the following properties,
\begin{equation}
\ve{G}(t+\pi,t'+\pi) = \ve{G}(t,t') \ (t \ge t') \ ,
\end{equation}
\begin{equation}
\ve{G}(t,t'')\ve{G}(t'',t') = \ve{G}(t,t') \ (t \ge t'' \ge t')
\end{equation}
and 
\begin{equation}
\ve{G}(t',t) = {\ve{G}(t,t')}^{-1} \ (t \ge t') \ .
\end{equation}
Therefore, for a natural number $N$, we obtain
\begin{eqnarray}
\ve{G}(t,t-N\pi) \aq \ve{G}(t,t-\pi)\ve{G}(t-\pi,t-2\pi)\times\cdots\times\ve{G}(t-(N-1)\pi,t-N\pi) \nonumber \\
 \aq \ve{G}(t,t-\pi)^N \nonumber \\
 \aq \left\{\ve{G}(t-\pi,0)\ve{G}(\pi,0)\ve{G}(t-\pi,0)^{-1}\right\}^N \nonumber \\
 \aq \ve{G}(t-\pi,0)\ve{G}(\pi,0)^N\ve{G}(t-\pi,0)^{-1} \ .
\label{eq4}
\end{eqnarray}

Now, we examine variances in the long-time limit which are obtained by setting $t_0=t-N\pi$ in Eq.(\ref{eq3}) and then letting $N$ goes to $\infty$. If the absolute values of all eigenvalues of $\ve{G}(\pi,0)$ are lower than 1 which is equivalent to the homogeneous solution being stable in the long-time limit, the first term of Eq.(\ref{eq3}) vanishes in such a limit because of Eq.(\ref{eq4}). We can find that the second term is also finite only if the eigenvalues fulfill the same condition as follows:
\begin{eqnarray}
\int_{t-N\pi}^{t}d t_1\ve{G}(t,t_1)\ve{K} \aq \sum_{n=0}^{N-1}\int_{t-(n+1)\pi}^{t-n\pi}d t_1\ve{G}(t,t_1)\ve{K} \nonumber \\
 \aq \sum_{n=0}^{N-1}\int_{t-\pi}^{t}d t_1\ve{G}(t,t_1-n\pi)\ve{K} \nonumber \\
 \aq \sum_{n=0}^{N-1}\ve{G}(t,t-\pi)^n\int_{t-\pi}^{t}d t_1\ve{G}(t,t_1)\ve{K} \nonumber \\
 &\longrightarrow& \left\{\ve{1} - \ve{G}(t,t-\pi)\right\}^{-1}\int_{t-\pi}^{t}d t_1\ve{G}(t,t_1)\ve{K} \ \ \ (N\longrightarrow\infty) \ .
\end{eqnarray}
In order that this sum of the matrix series converges, it is necessary that the absolute values of all eigenvalues of $\ve{G}(t,t-\pi)$ must be lower than 1. By setting $N=1$ in Eq.(\ref{eq4}), one can see that the eigenvalues of $\ve{G}(t,t-\pi)$ and $\ve{G}(\pi,0)$ are equal. If the absolute values of all eigenvalues of $\ve{G}(\pi,0)$ are lower than 1, this necessary condition is satisfied.

Then, variances in the long-time limit are given by
\begin{equation}
\ve{X}(t) = \left\{\ve{1} - \ve{G}(t,t-\pi)\right\}^{-1}\int_{t-\pi}^{t}d t_1\ve{G}(t,t_1)\ve{K} \ .
\label{eq15}
\end{equation}
By setting $t\mapsto t+\pi$ on the above equation one can obtain
\begin{equation}
\ve{X}(t+\pi) = \left\{\ve{1} - \ve{G}(t+\pi,t)\right\}^{-1}\int_{t}^{t+\pi}d t_1\ve{G}(t+\pi,t_1)\ve{K} = \left\{\ve{1} - \ve{G}(t,t-\pi)\right\}^{-1}\int_{t-\pi}^{t}d t_1\ve{G}(t+\pi,t_1+\pi)\ve{K} = \ve{X}(t) \ ,
\end{equation}
which means that variances have a periodicity of $\pi$.

\section{asymptotic behavior of MSD}
\label{variance}

At the previous section, we showed three variances in the long-time limit by use of $\ve{G}$ like Eq.(\ref{eq15}). This representation is useful for demonstrating the periodicity of them, but is not practical when investigating the asymptotic behavior of MSD. Thus, we shall derive another representation.

By vanishing ${\sigma_{v^2}}(t)$ and ${\sigma_{xv}}(t)$ from Eq.(\ref{eq2}) with $w=0$, we can obtain
\begin{equation}
\frac{d^3{{\sigma}}_{x^2}(t)}{d t^3} + 3\beta\frac{d^2{{\sigma}}_{x^2}(t)}{d t^2} + 2\left\{2q\phi(t)+\beta^2\right\}\frac{d{{\sigma}}_{x^2}(t)}{d t} + 2q\left\{2\beta\phi(t)+\frac{d{\phi}(t)}{d t}\right\}\sigma_{x^2}(t) = 4\epsilon \ .
\label{eq5}
\end{equation}
Since $\sigma_{x^2}(t)$ in the long-time limit is a periodic function of $\pi$, we expand it in a Fourier series
\begin{equation}
\sigma_{x^2}(t) = \sum_{n=-\infty}^{\infty}d_ne^{\is2nt} \ \ \ (d_n = {d_{-n}}^*) \ .
\label{eq6}
\end{equation}
We also expand $\phi(t)$ in a Fourier series
\begin{equation}
\phi(t) = \sum_{m=-\infty}^{\infty}c_me^{\is2mt} \ \ \ (c_m = {c_{-m}}^*) \ .
\label{eq7}
\end{equation}
By putting Eq.(\ref{eq6}) and Eq.(\ref{eq7}) into Eq.(\ref{eq5}) and arranging it, we can obtain
\begin{equation}
\sum_{n}\left[4\iu n(\beta+\iu n)(\beta+\iu2n)d_n + 4q\sum_{m}\left\{\iu(2n-m) + \beta\right\}d_{n-m}c_{m}\right]e^{\is2nt} = 4\epsilon \ .
\end{equation}
This equation can be expressed as 
\begin{equation}
\Lambda_{n}d_n + 4q\sum_{m}\left\{\iu(2n-m) + \beta\right\}c_{m}d_{n-m} = 4\epsilon\delta_{n,0} \ \ (n = 0,\pm1,\pm2,\cdots)
\label{eq8}
\end{equation}
where
\begin{equation}
\Lambda_{n} \equiv 4\iu n(\beta+\iu n)(\beta+\iu2n)
\label{eqL}
\end{equation}
and $\delta_{i,j}$ is the Kronecker symbol.

Let us represent Eq.(\ref{eq8}) by a matrix formulation
\begin{equation}
\left(\ve{\cal A} + q\ve{\cal B}\right)\ve{d} = \ve{\cal K}
\end{equation}
where
\begin{equation}
\ve{\cal A} \hspace{1ex} \equiv 
\begin{array}{r@{}l}
 \begin{array}{ccccc}
\hspace{-3.6cm} \cdots &\makebox[.4em]{}\scriptstyle -1 &\makebox[0.7em]{}\scriptstyle 0 &\makebox[.75em]{}\scriptstyle 1 & \hspace{.4em} \cdots
 \end{array} \\
\left(
\begin{array}{ccccc}
\ddots & \vdots & \vdots & \vdots & \iddots \\
\cdots & \Lambda_{-1} & 0 & 0 & \cdots \\
\cdots & 0 & 0 & 0 & \cdots \\
\cdots & 0 & 0 & \Lambda_1 & \cdots \\
\iddots & \vdots & \vdots & \vdots & \ddots
\end{array}
\right) & \begin{array}{r}
\raisebox{-0.0mm}{\vdots} \\ \raisebox{0.0em}{$\scriptstyle -1$} \\ \raisebox{0.ex}{$\scriptstyle0$} \\ \raisebox{0.mm}{$\scriptstyle1$} \\ \raisebox{0.mm}{\vdots}
\end{array}
\end{array} \ , \ \ 
\ve{\cal B} \hspace{1ex} \equiv 
\begin{array}{r@{}l}
 \begin{array}{ccccc}
\hspace{-7.3cm} \cdots &\makebox[1.7em]{}\scriptstyle -1 &\makebox[4.3em]{}\scriptstyle 0 &\makebox[4.5em]{}\scriptstyle 1 & \hspace{2.2em} \cdots
 \end{array} \\
\left(
\begin{array}{ccccc}
\ddots & \vdots & \vdots & \vdots & \iddots \\
\cdots & 4(\beta-\iu2)c_0 & 4(\beta-\iu)c_{-1} & 4\beta c_{-2} & \cdots \\
\cdots & 4(\beta-\iu)c_{1} & 4\beta c_0 & 4(\beta+\iu)c_{-1} & \cdots \\
\cdots & 4\beta c_2 & 4(\beta+\iu)c_1 & 4(\beta+\iu2)c_0 & \cdots \\
\iddots & \vdots & \vdots & \vdots & \ddots
\end{array}
\right) & \begin{array}{r}
\raisebox{-0.mm}{\vdots} \\ \raisebox{-0.em}{$\scriptstyle -1$} \\ \raisebox{-0.ex}{$\scriptstyle0$} \\ \raisebox{-0.mm}{$\scriptstyle1$} \\ \raisebox{-0.mm}{\vdots}
\end{array}
\end{array} \ ,
\label{eqB}
\end{equation}
\begin{equation}
\ve{d} \hspace{1ex} \equiv 
\begin{array}{r@{}l}
\left(
\begin{array}{c}
\vdots \\
d_{-1} \\
d_0 \\
d_1 \\
\vdots
\end{array}
\right) & \begin{array}{r}
\vdots \\ \raisebox{0.ex}{$\scriptstyle-1$} \\ \raisebox{0.ex}{$\scriptstyle0$} \\ \raisebox{0ex}{$\scriptstyle1$} \\ \vdots
\end{array}
\end{array} \ {\rm and} \ \ve{\cal K} \hspace{1ex} \equiv 
\begin{array}{r@{}l}
\left(
\begin{array}{c}
\vdots \\
0 \\
4\epsilon\\
0 \\
\vdots
\end{array}
\right) & \begin{array}{r}
 \vdots \\ \raisebox{0.ex}{$\scriptstyle-1$} \\ \raisebox{0.ex}{$\scriptstyle0$} \\ \raisebox{0ex}{$\scriptstyle1$} \\ \vdots
\end{array}
\end{array} \ .
\end{equation}
We number rows and columns of these matrices and vectors as shown at the sides of them.

For simplicity, we shall introduce a matrix $\ve{\cal D}(q)$ given by
\begin{equation}
\ve{\cal D}(q) \equiv \ve{\cal A} + q\ve{\cal B} \ .
\end{equation}
Hence, $\ve{d}$ can be solved like this:
\begin{equation}
\ve{d} = \ve{\cal D}(q)^{-1}\ve{K} \ .
\end{equation}
Furthermore we denote the determinant and the adjugate matrix of $\ve{\cal D}(q)$ by $\Delta(q)$ and $\hat{\ve{\cal D}}(q)$, respectively.

Here, let us estimate $\sigma_{x^2}(t)$ in the long-time limit by averaging over the period as past studies \cite{tashiro07,arnold93,hanggi}:
\begin{equation}
\overline{\sigma_{x^2}} \equiv \frac{1}{\pi}\int_{0}^{\pi}d t\sigma_{x^2}(t) = d_0(q) = 4\epsilon\left[{\ve{\cal D}(q)}^{-1}\right]_{0,0} \ .
\label{eq9}
\end{equation}
By using the determinant and the adjugate matrix, the above equation becomes
\begin{equation}
\overline{\sigma_{x^2}} = \frac{4\epsilon\hat{\cal D}_{0,0}(q)}{\Delta(q)} \ .
\end{equation}
When $q$ goes to 0, $\Delta(q)$ becomes 0 since all the elements of the zeroth row and the zeroth column of $\ve{\cal D}(0)$ are 0, which causes the divergence of MSD in the long-time limit. Of course, in order to reach the conclusion it is necessary that $\hat{\cal D}_{0,0}(0) \neq0$, but this becomes obvious later.

We assume that the order of the singular point is $n$, so that we can expand $\overline{\sigma_{x^2}}$ around $q=0$ as follows:
\begin{equation}
\overline{\sigma_{x^2}} = \frac{\alpha_{-n}}{q^n} + \frac{\alpha_{-n+1}}{q^{n-1}} + \cdots + \alpha_0 + \alpha_1q + \cdots \ .
\end{equation}
It is clear that the coefficient of $q^{-n}$ can be obtained by
\begin{equation}
\alpha_{-n} = 4\epsilon\lim_{q\rightarrow0}\frac{q^n}{\Delta(q)}\hat{\cal D}_{0,0}(q) \ .
\end{equation}

If the order is 1, by use of  L'Hospital's rule on the above equation $\alpha_{-1}$ becomes
\begin{equation}
\alpha_{-1} = 4\epsilon\frac{\hat{\cal D}_{0,0}(0)}{\dot{\Delta}(0)} = 4\epsilon\frac{\hat{\cal A}_{0,0}}{\dot{\Delta}(0)} \ ,
\label{alpha}
\end{equation}
in which the over-dot means the derivative with respect to $q$.

Note that $\dot{\Delta}(0)$ is not 0 if $c_0\ne0$. This can be understood in the following ways:
As is well known,
\begin{equation}
\dot{\Delta}(q) = \mbox{Tr}\left[\hat{\ve{\cal D}}(q)\dot{\ve{\cal D}}(q)\right] = \mbox{Tr}\left[\hat{\ve{\cal D}}(q)\ve{\cal B}\right] \ ,
\label{eq10}
\end{equation}
which leads to
\begin{equation}
\dot{\Delta}(0) = \mbox{Tr}\left[\hat{\ve{\cal D}}(0)\ve{\cal B}\right] = \mbox{Tr}\left[\hat{\ve{\cal A}}\ve{\cal B}\right] = \sum_{m,n}\hat{\cal A}_{n,m}{\cal B}_{m,n} \ ,
\label{eqZ}
\end{equation}
where $\hat{\ve{\cal A}}$ is the adjugate matrix of $\ve{\cal A}$.
From Eq.(\ref{eqB}), one can confirm that all the elements of the zeroth row and the zeroth column of $\ve{\cal A}$ are 0. Therefore, the adjugate matrix has only one nonzero element $\hat{{\cal A}}_{0,0} [=\hat{\cal D}_{0,0}(0)]$, i.e.,
\begin{equation}
\hat{\cal A}_{n,m} = \hat{\cal A}_{0,0}\delta_{n,0}\delta_{m,0} \ .
\end{equation}
Substituting this into Eq.(\ref{eqZ}), we obtain
\begin{equation}
\dot{\Delta}(0) = \hat{\cal A}_{0,0}{\cal B}_{0,0} = 4\beta c_0\hat{\cal A}_{0,0} \ .
\end{equation}

From the above equation, we can get
\begin{equation}
\alpha_{-1} = \frac{\epsilon}{\beta c_0} \ .
\label{coeff-1}
\end{equation}
If $c_0>0$, $\overline{\sigma_{x^2}}$ is positive for small $q$. On the other hand, if $c_0<0$, $\overline{\sigma_{x^2}}$ becomes negative, which cannot be understood physically. This contradiction implies that the area where $w=0$ and $q$ is small belongs to an {\em unstable region}, which is made up of parameters which make this system unstable in the long-time limit. Thus, $\overline{\sigma_{x^2}}$ cannot be defined. As an example, we have shown stable and unstable regions of the system where $\phi(t)$ is a square wave in the past paper \cite{tashiro06}.  From Fig.2(c) in Ref.\cite{tashiro06} corresponding to the case with  $c_0<0$, one can see that an unstable region includes a line $w=0$ within small $q$.

If $c_0=0$, $\alpha_{-1}$ diverges to $\infty$. Thus, the order of the singular point is more than 2. If we assume that the order is 2, $\alpha_{-2}$ can be derived as
\begin{equation}
\alpha_{-2} = 8\epsilon\frac{\hat{\cal A}_{0,0}}{\ddot{\Delta}(0)} \ .
\label{coeff-2}
\end{equation}
In this derivation, we have used L'Hospital's rule.
By differentiating Eq.(\ref{eq10}) with respect to $q$ and setting $q=0$, one can obtain
\begin{equation}
\ddot{\Delta}(0) = \mbox{Tr}\left[\dot{\hat{\ve{\cal D}}}(0)\ve{\cal B}\right] \ .
\label{ddotlambda}
\end{equation}
The elements of $\dot{\hat{\ve{\cal D}}}(0)$
can be calculated as follows:
\begin{equation}
\dot{\hat{\cal D}}_{n,m}(0) = \left\{
\begin{array}{l}
\displaystyle-\frac{\hat{\cal A}_{0,0}{\cal B}_{0,m}}{\Lambda_m}\delta_{n,0} \ \ \ (m\neq 0) \\
 \\
\displaystyle-\frac{\hat{\cal A}_{0,0}{\cal B}_{n,0}}{\Lambda_n}\delta_{m,0} \ \ \ (n\neq 0) \\
\end{array}
\right. \ .
\end{equation}
(See appendix \ref{app4}.)
It does not matter that $\dot{\hat{\cal D}}_{0,0}(0)$ is not determined by the above equation, since ${\cal B}_{0,0}=0$.
Then, equation (\ref{ddotlambda}) with the above results becomes
\begin{eqnarray}
\ddot{\Delta}(0) \aq \sum_{m,n}\dot{\hat{\cal D}}_{n,m}(0){\cal B}_{m,n} \nonumber \\
 \aq \sum_{m\neq0}\dot{\hat{\cal D}}_{0,m}(0){\cal B}_{m,0} + \sum_{n\neq0}\dot{\hat{\cal D}}_{n,0}(0){\cal B}_{0,n} \nonumber \\
 \aq -\hat{\cal A}_{0,0}\left(\sum_{m\neq0}\frac{{\cal B}_{0,m}{\cal B}_{m,0}}{\Lambda_m}+\sum_{n\neq0}\frac{{\cal B}_{n,0}{\cal B}_{0,n}}{\Lambda_n}\right) = -2\hat{\cal A}_{0,0}\sum_{m\neq0}\frac{{\cal B}_{0,m}{\cal B}_{m,0}}{\Lambda_m} \ .
\label{ddotlambda2}
\end{eqnarray}
Moreover, from Eq.(\ref{eqB}) it is easily confirmed that the elements of the zeroth row and the zeroth column of $\ve{\cal B} $ are 
\begin{equation}
{\cal B}_{0,m} = 4(\beta+\iu m)c_{-m} \ \mbox{and} \ {\cal B}_{m,0} = 4(\beta+\iu m)c_{m} \ .
\end{equation}
By inserting these representations into Eq.(\ref{ddotlambda2}), $\ddot{\Delta}(0)$ becomes
\begin{equation}
\ddot{\Delta}(0) = -2\hat{\cal A}_{0,0}\sum_{m\neq0}\frac{4(\beta+\iu m)|c_m|^2}{\iu m(\beta+\iu2m)} = 16\hat{\cal A}_{0,0}\sum_{m=1}^{\infty}\frac{\beta|c_m|^2}{\beta^2+4m^2} \ ,
\end{equation}
which yields
\begin{equation}
\alpha_{-2} = \frac{\epsilon}{\sum_{m=1}^{\infty}\frac{2\beta|c_m|^2}{\beta^2+4m^2}} \ .
\end{equation}
This means that the order with the case $c_0=0$ is 2.

We can reach the following results:
\begin{equation}
\overline{\sigma_{x^2}} = \left\{
\begin{array}{cc}
\displaystyle \frac{\epsilon}{\beta}\frac{1}{\sum_{m=1}^{\infty}\frac{2|c_m|^2}{\beta^2+4m^2}}\frac{1}{q^2} + O(q^{-1}) & (c_0=0) \\
 & \\
\displaystyle\frac{\epsilon}{\beta}\frac{1}{c_0}\frac{1}{q} + O(q^0) & (c_0>0)
\end{array}
\right. \ .
\label{eqMainResult}
\end{equation}
Note that the order of the singular point depends on whether $c_0$ is 0 or not.

For the Mathieu's case of $\phi(t) = \cos2t$ where $|c_m|^2=\delta_{|m|,1}/4$, the equation (\ref{eqMainResult}) becomes
\begin{equation}
\overline{\sigma_{x^2}} = \frac{\epsilon}{\beta}\frac{2(\beta^2+4)}{q^2} + O(q^0) \ .
\end{equation}
This result corresponds to the past works\cite{hanggi,arnold93} 
by setting $q\longmapsto2q$.

By these results, we can know the asymptotic behavior of MSD in the long-time limit with a general periodic function.
These expressions including only one term are useful for finding a periodic function which reduces MSD. By comparing the coefficient of $q$ or $q^2$, we can find such a periodic function more smoothly than calculating MSD haphazardly.

\section{periodic functions reducing MSD}

Let us find a periodic function which makes MSD in the long-time limit smaller than that for the Mathieu's case of $\phi(t) = \cos2t$ in the following two cases: $c_0=0$ and $c_0>0$. We set its unit $\epsilon/\beta$ when showing MSD.

Here,  we review how $w$, $q$ and $\phi(t)$ are determined which is explained in Ref.\cite{tashiro07}. $w$ expresses the middle value between the maximum and the minimum of an arbitrary periodic function with period $\pi$. $q$ denotes the interval between the maximum and $w$ (between the minimum and $w$).
Then, this periodic function is expressed as $w+q\phi(t)$. Therefore, the period and the amplitude of $\phi(t)$ must be $\pi$ and 1, respectively. Needless to say, $\cos2t$ satisfies these conditions. We will compare MSD for the Mathieu's case with that for a periodic function satisfying them.

\subsection{periodic function with $c_0=0$}

\begin{figure}[ht]
\begin{center}
 \includegraphics[scale=1.]{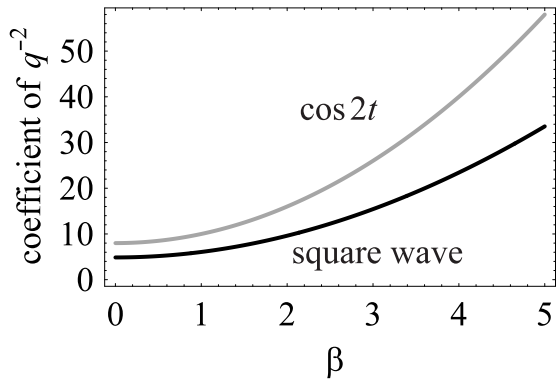}
 \caption{Comparison of the coefficients of $q^{-2}$ between a square wave (black curve) and cosine (dark curve). The coefficients are plotted as functions of $\beta$.}
 \label{fig1}
\end{center}
\end{figure}

We propose a square wave
\begin{equation}
\phi(t) = \left\{
\begin{array}{rcc}
1& \mbox{if} & (n-1)\pi\le t\le (n-\frac{1}{2})\pi \ , \\
-1 & \mbox{if} & (n-\frac{1}{2})\pi\le t\le n\pi \ ,
\end{array}
\right. \ (n=0,\pm1,\pm2,\cdots)
\end{equation}
as a periodic function whose $c_0$ is equal to 0. The square of the absolute value of coefficients is
\begin{equation}
|c_n|^2 = \frac{2\{1-(-1)^n\}}{n^2\pi^2} \ .
\end{equation}
With this $|c_n|^2$, the coefficient of $q^{-2}$ is derived as
\begin{equation}
\frac{1}{\sum_{m=1}^{\infty}\frac{2|c_m|^2}{\beta^2+4m^2}} = \frac{\pi\beta^3}{\pi\beta-4\tanh(\pi\beta/4)} \ .
\end{equation}
This coefficient of $q^{-2}$ is smaller than that of the Mathieu's case. We plot two coefficients in Fig.\ref{fig1}, from which one can see that the coefficient of a square wave is about half that of the Mathieu's case. Therefore, we can presume that MSD for a square wave must be smaller than that for the Mathieu's case.

\begin{figure}[ht]
\begin{center}
  \includegraphics[scale=1.2]{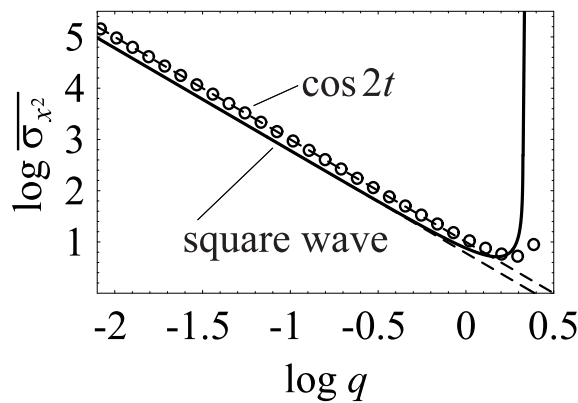}
   \caption{$\overline{\sigma_{x^2}}$ at $\beta=1$ as a function of $q$. The circles denote the Mathieu's case ($\cos2t$) derived by solving Eq.(\ref{eq2}) numerically. The solid curve means the analytical solution for a square wave. The two dashed lines are asymptotic solutions [Eq.(\ref{eqMainResult})] for each case.}
 \label{fig11}
\end{center}
\end{figure}

In Fig.\ref{fig11}, $\overline{\sigma_{x^2}}$ at $\beta = 1$ for a square wave and the Mathieu's case are shown. The circles are numerical solutions for the Mathieu's case. The solid curve is a solution for a square wave, which we can derive analytically \cite{tashiro06}. The two dashed lines mean asymptotic solutions, Eq.(\ref{eqMainResult}), for each case. As one can see, both asymptotic solutions correspond to the numerical and the analytical solution at a region where $q$ is small, which guarantees that the result in the previous section is valid.

\begin{figure}[ht]
\begin{center}
 \includegraphics[scale=1.2]{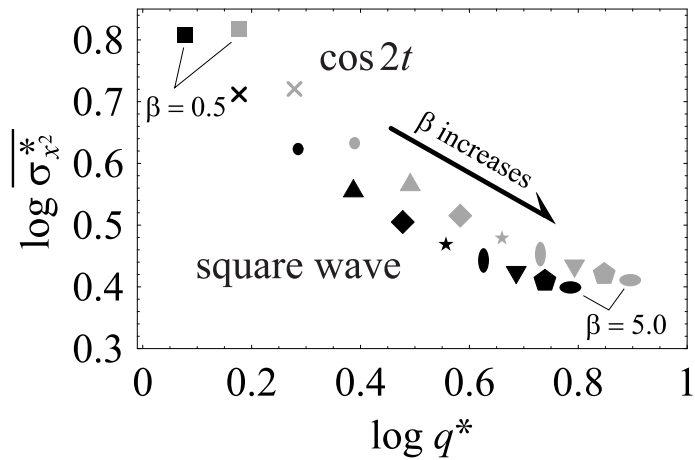}
 \caption{$\overline{\sigma_{x^2}^{*}}$ which represents a minimum value of $\overline{\sigma_{x^2}}$ vs $q^*$ which minimizes $\overline{\sigma_{x^2}}$. Gray and black symbols represent values for the Mathieu's case ($\cos2t$) and a square wave, reactively. The same symbol refers to the same $\beta$. As the symbol changes from the upper left to the lower right on this figure, $\beta$ gets larger from 0.5 to 5.0 in steps of 0.5.}
 \label{fig21}
\end{center}
\end{figure}

As we presumed, MSD for a square wave is smaller.  However, the two minimum values seem to be close from this figure. From numerical computations, it turns out that the minimum for a square wave is 5.15 at $q=1.50$ and that for the Mathieu's case is 5.25 at $q=1.90$. Therefore, the minimum is also smaller.
In  Fig.\ref{fig21}, we show the minimum value of $\overline{\sigma_{x^2}}$, which we denote by $\overline{\sigma_{x^2}^{*}}$ and $q^*$ which minimizes $\overline{\sigma_{x^2}}$ at several values of $\beta$. The same symbol refers to the same $\beta$. From the upper left to the lower right on this figure, $\beta$ increases from 0.5 to 5.0 in steps of 0.5. For this range of $\beta$, the minimum values for a square wave are always lower than those for the Mathieu's case. By setting $\log q\approx 0.4$, however, $\overline{\sigma_{x^2}}$ for the Mathieu's case can be lower than that for a square wave as seen in Fig.\ref{fig11}, which arises from the fact that the range of $q$ which makes this system stable for a square wave is narrower than for the Mathieu's case.

\subsection{periodic function with $c_0>0$}

We define a periodic function with $c_0>0$ in the first interval from $0$ to $\pi$ as follows: First, we form the positive part of this periodic function at $0\le t\le T_1$ from $\psi(t/T_1)$ where $T_1<\pi$ and $\psi(t)$ is a non-negative function defined on $0\le t\le 1$ whose maximum value is 1. Next, we form the negative part at $T_1\le t\le \pi$ from $-\psi((t-T_1)/T_2)$ with $T_2\equiv\pi-T_1$. Then we join these two parts together, so that we can get the periodic function in the interval like this: $\psi(t/T_1)\left\{1-\Theta(t-T_1)\right\} - \psi((t-T_1)/T_2)\Theta(t-T_1) \ \ (0\le t\le \pi)$ in which $\Theta(t)$ is the Heaviside's step function.
 Moreover, the $\pi$-periodic function fulfilling the conditions mentioned at the beginning of this section is completed by connecting this segment repeatedly.

Using the periodic function defined in this way, we derive
\begin{equation}
c_0 = \frac{1}{\pi}\int_0^{T_1}dt\psi(t/T_1) - \frac{1}{\pi}\int_0^{T_2}dt\psi(t/T_2) = \frac{T_1-T_2}{\pi}\int_0^{1}dt\psi(t) = \frac{2T_1-\pi}{\pi}\int_0^{1}dt\psi(t) \ .
\end{equation}
Here, we introduce the ratio of the time intervals of the two parts, $\gamma\equiv T_2/T_1$, denoting the asymmetry of this periodic function. Then, $c_0$ becomes
\begin{equation}
c_0 = \frac{1-\gamma}{1+\gamma}\int_0^{1}dt\psi(t) \ ,
\end{equation}
because $T_1=\frac{\pi}{1+\gamma}$.
Since $c_0$ must be positive, it is necessary that $0<\gamma<1$. From this equation, we can find that as $\gamma$ decreases with fixing $\psi(t)$, which indicates that the asymmetry becomes extreme, $c_0$ becomes larger, which means that MSD becomes reduced because of Eq.(\ref{eqMainResult}). Therefore, we can reach a conclusion that the asymmetry suppresses the fluctuation of position for the case $w=0$ as well as the case $w>0$ reported in Ref.\cite{tashiro07}.

\begin{figure}[ht]
\begin{center}
 \includegraphics[scale=0.8]{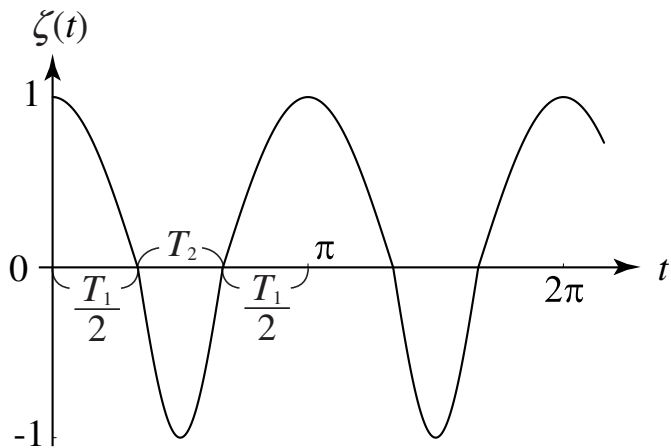}
 \caption{An asymmetric cosine function $\zeta(t)$. The time intervals where $\zeta$ is positive ($T_1$) and $\zeta$ is negative ($T_2$) are not equal.}
 \label{fig2}
\end{center}
\end{figure}

As an example of such a periodic function we show 
an asymmetric cosine function $\zeta(t)$ (see Fig.\ref{fig2}) which can be obtained by setting $\psi(t)=\sin\pi t$ and inserting the negative part into the positive part divided equally. Of course, this procedure does not affect the above discussion. Thus, we can get
\begin{equation}
c_0 = \frac{2}{\pi}\frac{1-\gamma}{1+\gamma} \ .
\label{eqc0}
\end{equation}
Moreover, we find
\begin{equation}
c_n = e^{\is nT_1}\left\{\frac{(1+e^{-\is2nT_1})T_1}{\pi^2-4n^2T_1^2} - \frac{(1+e^{\is2n\gamma T_1})\gamma T_1}{\pi^2-4n^2\gamma^2T_1^2}\right\} \ \  (n = \pm1, \pm2, \cdots) \ .
\end{equation}

\begin{figure}[ht]
\begin{center}
 \includegraphics[scale=1.2]{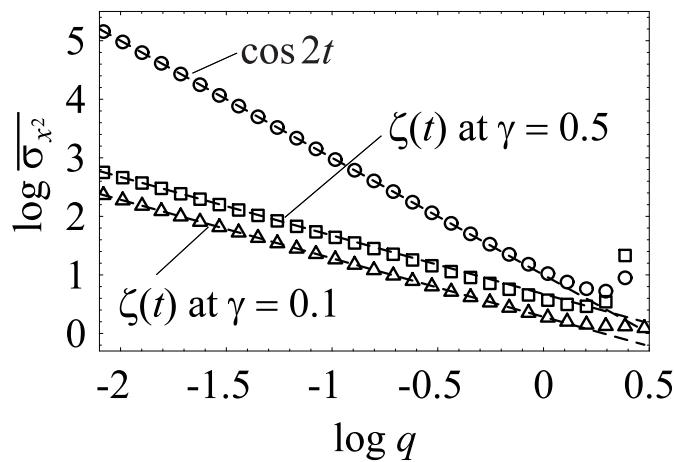}
 \caption{$\overline{\sigma_{x^2}}$ as a function of $q$ at $\beta=1$. The triangles and the boxes represent values derived from Eq.(\ref{eq2}) numerically for an asymmetric cosine $\zeta(t)$ at $\gamma=0.1$ and $\gamma=0.5$, respectively. The circles represent the Mathieu's case ($\cos2t$). Dashed lines are asymptotic solutions described by Eq.(\ref{eqMainResult}) for each case.}
 \label{fig12}
\end{center}
\end{figure}

Figure \ref{fig12} shows $\overline{\sigma_{x^2}}$ at $\beta=1$ as a function of $q$. The triangles and the boxes represent numerical solutions for an asymmetric cosine at $\gamma=0.1$ and $\gamma=0.5$, respectively. The circles represent numerical solutions for the Mathieu's case. Dashed lines represent asymptotic solutions described by Eq.(\ref{eqMainResult}) for each case which coincide with numerical solutions at small $q$. 
Since the orders of the singular point are different between the two cases, these lines are crossing. 

For small $q$, the difference between MSD for an asymmetric cosine and for the Mathieu's case is extreme.
MSD at $\gamma=0.1$ is smaller than that at $\gamma=0.5$, which can be easily understood as follows: From Eq.(\ref{eqc0}), $c_0$ decreases monotonically as $\gamma$ increases. The coefficient of $q^{-1}$ is proportional to $1/c_0$, and so MSD increases when $\gamma$ increases. For small $q$, the ratio of MSD at $\gamma=0.1$ to at $\gamma=0.5$ is about 40.7\%. The minimum value of MSD for the Mathieu's case is larger than that for  $\gamma=0.1$ and  $\gamma=0.5$. Even if $\gamma$ gets bigger, the large-and-small relation does not change: from numerical calculations, it is clarified that although $\gamma$ varies up to 0.9 the minimum for an asymmetric cosine does not become more than that for the Mathieu's case at the range of $\beta$ from 0.1 to 5.0.

\section{concluding remarks}

In this paper, we considered analytically the Brownian parametric oscillator described by Eq.(\ref{eq1}) especially with setting $w=0$ which is not treated in \cite{tashiro07} and the following points are made clear: I. It is proved that the period of the variances becomes the same as that of $\phi(t)$ in the long-time limit. II. asymptotic behaviors are derived as Eq.(\ref{eqMainResult}), which are the main results of this paper and are applicable to general periodic functions.  III. A square wave and an asymmetric cosine are proposed as periodic functions with $c_0=0$ and $c_0>0$, respectively, which make MSD smaller than that for the Mathieu's case.

One of main results, $\overline{\sigma_{x^2}} = \epsilon/(\beta c_0q) + O(q^0)$, can be obtained more easily from another point of view: If $q$ is very small, the motion can be separated into a ``fast'' part whose period is $\pi$ and a ``slow'' part which hardly changes during the period \cite{landau}. By taking the time average of the total motion, the fast part vanishes. The effective potential which induces the slow part is written as \cite{adia}
\begin{equation}
\frac{1}{2}\left(qc_0 + \frac{q^2}{2}\sum_{m=1}^{\infty}\frac{|c_m|^2}{m^2}\right)\overline{x(t)}^2 \ ,
\end{equation}
where $\overline{x(t)}$ denotes the slow part of the motion. Hence this slow part is a Brownian motion in the harmonic potential and its MSD in the long-time limit can be calculated like
\begin{equation}
\frac{\epsilon}{\beta}\frac{1}{qc_0 + \frac{q^2}{2}\sum_{m=1}^{\infty}\frac{|c_m|^2}{m^2}}
\label{eqadia}
\end{equation}
whose first term (proportional to $q^{-1}$) is compatible with our result.
On the other hand, equation (\ref{eqadia}) with $c_0=0$ does not correspond to our result \footnote{If we sum up all higher order terms which are not derived in Ref.\cite{adia}, the two results may agree.}. If we assume that $\beta$ is small, however, they become equal. Therefore, it is the achievement of this paper that the asymptotic solution for $c_0=0$ can be acquired precisely.

Our results indicate that MSD can be reduced more remarkably than ever in experiments where this Brownian motion can be observed, e.g., experiments using the Paul trap with a buffer gas \cite{arnold93}.
The author hopes that this result will be beneficial in such experiments.

\section{acknowledgment}

The author would like to thank Dr. Akio Morita for the extensive discussions.

\appendix

\section{differential equation governing three variances}
\label{app3}

The derivatives of three variances shown in Eq.(\ref{vars}) are
\begin{equation}
\frac{{d}}{d t}\left(
\begin{array}{c}
\av{\sigma_{x^2}(t)} \\
\av{\sigma_{v^2}(t)} \\
\av{\sigma_{xv}(t)}
\end{array}
\right) = \left(
\begin{array}{c}
2\sigma_{xv}(t) \\
2\av{v(t)\dot{v}(t)} \\
\sigma_{v^2}(t) + \av{x(t)\dot{v}(t)}
\end{array}
\right) \ .
\label{eqA-4}
\end{equation}
From Eq.(\ref{eq1}), we can obtain
\begin{equation}
\av{v(t)\dot{v}(t)} = -\beta\av{v(t)^2} - [w + q\phi(t)]\av{x(t)v(t)} + \av{v(t)f(t)}
\end{equation}
and
\begin{equation}
\av{x(t)\dot{v}(t)} =
-\beta\av{x(t)v(t)} - [w + q\phi(t)]\av{x(t)^2} + \av{x(t)f(t)} \ .
\label{eqA-5}
\end{equation}
$\av{v(t)f(t)}$ and $\av{x(t)f(t)}$ are derived by use of Novikov's theorem \cite{nov} which states that
a functional of Gaussian noise $g[f(t)]$ satisfies
\[
\av{g[f(t)]f(t)} = \int_{0}^{t}d t'\av{f(t)f(t')}\left.\av{\frac{\delta g[f(t)]}{\delta f(t')}}\right|_{t=t'}
\]
where ${\delta g[f(t)]}/{\delta f(t')}$ indicates the functional derivative of ${g[f(t)]}$ with respect to $f(t')$.

Because of
\begin{equation}
x(t) = x(0) + \int_{0}^{t}d t'v(t')
\end{equation}
and
\begin{equation}
v(t) = v(0) + \int_{0}^{t}d t'\left\{-\beta\dot{x}(t') - [w+q\phi(t')]x(t') + f(t')\right\} \ ,
\end{equation}
we get
\begin{equation}
\frac{\delta x(t')}{\delta f(t')} = 0 \ \ \mbox{and} \ \ \frac{\delta v(t')}{\delta f(t')} = 1 \ ,
\end{equation}
which yield
\begin{equation}
\av{x(t)f(t)} = 0 \ \ \mbox{and} \ \ \av{v(t)f(t)} = \epsilon \ .
\end{equation}
Therefore equation (\ref{eqA-4}) becomes 
\begin{equation}
\frac{{d}}{d t}\left(
\begin{array}{c}
\av{x(t)^2} \\
\av{v(t)^2} \\
\av{x(t)v(t)}
\end{array}
\right) = \left(
\begin{array}{ccc}
0 & 0       & 2 \\
0 & -2\beta  & -2[w + q\phi(t)] \\
-[w + q\phi(t)] & 1 & -\beta
\end{array}
\right)\left(
\begin{array}{c}
\av{x(t)^2} \\
\av{v(t)^2} \\
\av{x(t)v(t)}
\end{array}
\right) + \left(
\begin{array}{c}
0 \\
2\epsilon \\
0
\end{array}
\right) \ .
\label{eqA-6}
\end{equation}

\section{derivation of $\dot{\hat{\cal D}}_{n,m}(0)$}
\label{app4}

As is well known,
$\ve{\cal D}(q)$ and $\hat{\cal \ve{\cal D}}(q)$ satisfy the following relation
\begin{equation}
\ve{\cal D}(q)\hat{\cal \ve{\cal D}}(q) = \hat{\cal \ve{\cal D}}(q)\ve{\cal D}(q) = \Delta(q)\ve{E}
\end{equation}
where $\ve{E}$ is the infinite unit matrix. 
By differentiating this relation with respect to $q$ and setting $q=0$, one can obtain
\begin{equation}
\dot{\hat{\ve{\cal D}}}(0)\ve{\cal A} = -\hat{\ve{\cal A}}\ve{\cal B}
\label{eq9-6} \\
\end{equation}
and
\begin{equation}
\ve{\cal A}\dot{\hat{\ve{\cal D}}}(0) = -\ve{\cal B}\hat{\ve{\cal A}} \ . 
\label{eq9-7}
\end{equation}
In this derivation, we have used $\dot{\Delta}(0)=0$.

The elements of the $n$th row and the $m$th column of both hand sides of Eq.(\ref{eq9-6}) are
\begin{equation}
\sum_{l}\dot{\hat{\cal D}}_{n,l}(0){\cal A}_{l,m} = \Lambda_{m}\dot{\hat{\cal D}}_{n,m}(0)
\end{equation}
and
\begin{equation}
-\sum_{l}\hat{\cal A}_{n,l}{\cal B}_{l,m} = -\sum_{l}\hat{\cal A}_{0,0}\delta_{n,0}\delta_{l,0}{\cal B}_{l,m} = -\hat{\cal A}_{0,0}{\cal B}_{0,m}\delta_{n,0} \ .
\end{equation}
Keeping $\Lambda_0=0$ in mind, we can derive
\begin{equation}
\dot{\hat{\cal D}}_{n,m}(0) = -\frac{\hat{\cal A}_{0,0}{\cal B}_{0,m}}{\Lambda_m}\delta_{n,0} \ \ \ (m\neq 0) \ .
\end{equation}
Similarly, from Eq.(\ref{eq9-7}) we can also derive
\begin{equation}
\dot{\hat{\cal D}}_{n,m}(0) = -\frac{\hat{\cal A}_{0,0}{\cal B}_{n,0}}{\Lambda_n}\delta_{m,0} \ \ \ (n\neq 0) \ .
\end{equation}

\end{document}